\documentclass[aps,floats]{revtex4}
\usepackage{amsmath,amssymb}
\usepackage{graphicx,epsfig}
\usepackage[greek,english]{babel}

\begin{document}
\bibliographystyle {plain}

\def\oppropto{\mathop{\propto}} 
\def\opsimeq{\mathop{\simeq}}
\def\opoverderline{\mathop{\overline}}
\def\operarrow{\mathop{\longrightarrow}}
\def\opsim{\mathop{\sim}}

\def\fig#1#2{\includegraphics[height=#1]{#2}}
\def\figx#1#2{\includegraphics[width=#1]{#2}}


\title{ Level repulsion exponent $\beta$ for Many-Body Localization Transitions \\
and for Anderson Localization Transitions via  Dyson Brownian Motion} 


\author{ C\'ecile Monthus }
 \affiliation{Institut de Physique Th\'{e}orique, 
Universit\'e Paris Saclay, CNRS, CEA,
91191 Gif-sur-Yvette, France}

\begin{abstract}
The generalization of the Dyson Brownian Motion approach of random matrices to Anderson Localization (AL) models [Chalker, Lerner and Smith PRL 77, 554 (1996)] and to Many-Body Localization (MBL) Hamiltonians [Serbyn and Moore arxiv:1508.07293] is revisited to extract the level repulsion exponent $\beta$, where $\beta=1$ in the delocalized phase governed by the Wigner-Dyson statistics, $\beta=0$ in the localized phase governed by the Poisson statistics, and $0<\beta_c<1$ at the critical point. The idea is that the Gaussian disorder variables $h_i$ are promoted to Gaussian stationary processes $h_i(t)$ in order to sample the disorder stationary distribution with some time correlation $\tau$. The statistics of energy levels can be then studied via Langevin and Fokker-Planck equations. For the MBL quantum spin Hamiltonian with random fields $h_i$, we obtain $\beta =2q^{EA}_{n,n+1}(N)/q^{EA}_{n,n}(N)  $ in terms of the Edwards-Anderson matrix $q^{EA}_{nm}(N) \equiv \frac{1}{N} \sum_{i=1}^N  \vert < \phi_n \vert \sigma_i^z \vert \phi_m> \vert^2 $ for the same eigenstate $m=n$ and for consecutive eigenstates $m=n+1$.  For the Anderson Localization tight-binding Hamiltonian with random on-site energies $h_i$, we find $\beta =2 Y_{n,n+1}(N)/(Y_{n,n}(N)-Y_{n,n+1}(N)) $ in terms of  the Density Correlation matrix $Y_{nm}(N) \equiv  \sum_{i=1}^N  \vert < \phi_n \vert i> \vert^2 \vert <i \vert \phi_m> \vert^2 $ for consecutive eigenstates $m=n+1$, while the diagonal element $m=n$ corresponds to the Inverse Participation Ratio $Y_{nn}(N) \equiv  \sum_{i=1}^N  \vert < \phi_n \vert i> \vert^4 $ of the eigenstate $\vert \phi_n>$.

\end{abstract}

\maketitle

\section { Introduction}

The statistics of energy levels in quantum many-body systems has a long history
 since the pioneering work of Wigner \cite{wigner} and Dyson
\cite{dysonRMT} who introduced random matrices to understand the properties of spectra in nuclear physics \cite{mehta,oxford}. One essential property is the exponent $\beta$ governing the level repulsion
 in the probability distribution 
of the interval $s=E_{n+1}-E_n$ between two consecutive energy levels
\begin{eqnarray}
P(s) \oppropto_{s \to 0} s^{\beta}
\label{psbeta}
\end{eqnarray}
that directly 
reflects how much the corresponding eigenstates 'see' each other.
For real random matrices (GOE ensemble), the Wigner-Dyson statistics corresponds to the linear level repulsion \cite{mehta,oxford}
\begin{eqnarray}
\beta^{GOE}=1
\label{betagoe}
\end{eqnarray}
whereas other random ensembles lead to the other values $\beta=2$, $\beta=4$ \cite{mehta,oxford}.
The other famous universality class
 is the Poisson statistics of independent random energies
with no level repulsion
\begin{eqnarray}
\beta^{Poisson}=0
\label{betapoisson}
\end{eqnarray}
In the field of quantum chaos, the Poisson statistics
appears for systems whose classical dynamics is integrable \cite{berry},
whereas the Wigner-Dyson statistics appears when the
 classical dynamics is chaotic \cite{bohigas}.

In the field of Many-Body-Localization  (MBL) (see the recent reviews \cite{revue_huse,revue_altman} and references therein), the Wigner-Dyson statistics
appears in the delocalized phase where the Eigenstate Thermalization Hypothesis (E.T.H.) \cite{deutsch,srednicki,nature,mite,rigol} holds,
whereas the Poisson statistics appears in the Many-Body Localized phase,
which is characterized by an extensive number 
of emergent localized conserved operators \cite{emergent_swingle,emergent_serbyn,emergent_huse,emergent_ent,imbrie,serbyn_quench,emergent_vidal,emergent_ros,emergent_rademaker}. These conserved operators can be for instance constructed via the RSRG-X procedure \cite{rsrgx,rsrgx_moore,vasseur_rsrgx,yang_rsrgx,rsrgx_bifurcation,vasseur_edge,c_emergent} that generalize to the eigenstates the Fisher Strong Disorder Real Space RG for groundstates \cite{fisher_AF,fisher,fisherreview}. This RSRG-X approach breaks down as the MBL transition towards delocalization is approached as a consequence of resonances, and other types of RG have been proposed for the critical region \cite {vosk_rgentanglement,vasseur_resonant}.

The statistics of energy levels can be considered as the simplest criterion
to locate the Many-Body-Localization transition between these two phases 
\cite{pal,vadim,alet}, since the other criteria are based on the properties of
eigenfunctions, in particular on their entanglement properties \cite{bauer,kjall,alet,grover}. At criticality, one expects a non-trivial critical statistics
intermediate between Wigner-Dyson and Poisson, i.e. with an intermediate
level repulsion exponent 
\begin{eqnarray}
0< \beta^{criti} <1
\label{betacriti}
\end{eqnarray}
an intermediate level compressibility, and so on.
These critical statistics have been much studied for Anderson localization transitions, where they are related to the multifractal properties of eigenfunctions (see the reviews \cite{mirlinrevue} and references therein). In particular, many results have been obtained in the 'weak-multifractality regime' where the statistics is close to the Wigner Dyson  \cite{mirlinrevue}
and in the 'strong multifractality regime'  where the statistics is close to Poisson \cite{mirlin_evers,fyodorov,fyodorovrigorous,oleg1,oleg2,oleg3,oleg4,us_strongmultif,olivier_per,olivier_strong,olivier_conjecture}. 
For the Many-Body-Localization Transition, the singular 
perturbative approach around the Poisson limit has been studied in \cite{c_reso}.

To better understand the level statistics for Random Matrices, 
Dyson has introduced
the so-called Brownian motion model for the eigenvalues \cite{dyson}.
 Chalker, Lerner and Smith \cite{chalker} have adapted
 this approach to Anderson Localization models,
in order to relate the level statistics to the properties of eigenstates.
Very recently, Serbyn and Moore \cite{serbyn} have generalized
 this approach to Many-Body Localization models,
to understand the level statistics in terms of the properties of matrix elements of local operators, whose behavior is expected to change at the transition \cite{papic}.
In this paper we revisit this Brownian Dyson
 approach for Anderson Localization
and Many-Body localization Hamiltonians in order to extract the level repulsion exponent $\beta$.

The paper is organized as follows.
In section \ref{sec_dyson}, the Dyson Brownian approach is 
described for disordered Hamiltonians to obtain the Langevin equations 
for the rescaled energy levels in the middle of the spectrum.
In section \ref{sec_mbl}, the case of Many-Body-Localization quantum spin
Hamiltonian with random fields is analyzed in terms of Edwards-Anderson 
matrix elements. 
In section \ref{sec_anderson}, the case of Anderson Localization models
with random on-site energies is studied in terms of the Density correlation matrix
between eigenstates.
Our conclusions are summarized in section \ref{sec_conclusion},
and the two appendices contain some technical details.

\section{ Dyson Brownian motion for disordered Hamiltonians }

\label{sec_dyson}

\subsection{ Disordered Hamiltonian }

In this section, the Dyson approach is described for Hamiltonians
concerning $N$ sites of the form
\begin{eqnarray}
H = H_0 + \sum_{i=1}^N h_i O_i
\label{H}
\end{eqnarray}
 where $H_0$ is the non-random part, $O_i$ is a local operator on site $i$,
and the $h_i $ are independent Gaussian variables
of zero mean and variance  $W^2$ 
\begin{eqnarray}
 G_{W^2}(h_i) && = \frac{1}{\sqrt{2 \pi W^2} } e^{- \frac{h_i^2}{2 W^2}}
\label{gauss}
\end{eqnarray}
that define the disorder realization.

The form of Eq. \ref{H} is of course not restrictive, and one can adapt the method to other random terms like random couplings for instance, since the original Dyson approach is for fully random Hamiltonians, i.e. Random Matrices \cite{dyson}.

\subsection{ Random variables $h_i$ promoted to stochastic processes $h_i(t)$ }

The Dyson idea is to promote the random variables $h_i$
into independent stochastic processes $h_i(t)$ \cite{dyson,chalker,serbyn}.

Since one wishes to have the Gaussian distributions of Eq. \ref{gauss}
as stationary distributions, the simplest possibility is
to use Ornstein-Uhlenbeck processes, i.e. Gaussian stationary
Markov processes, characterized by some time correlation $\tau$
(see for instance the textbooks \cite{gardiner,vankampen,risken}).
They are defined by the Langevin equations
\begin{eqnarray}
\frac{d h_i(t)}{dt} && = -\frac{h_i(t)}{\tau}   + \xi_i(t)
\label{langevinhi}
\end{eqnarray}
with the linear restoring forces $(-h_i(t)/\tau)$ towards the origin and
with the independent white noises
\begin{eqnarray}
< \xi_{i}(t)  > && = 0
\nonumber \\
< \xi_{i}(t) \xi_{j}(t')) > && = 2 \frac{W^2}{\tau}  \delta(t-t') \delta_{ij}
\label{correbbxi}
\end{eqnarray}
Although present in the Dyson Brownian Motion paper \cite{dyson}, the restoring forces
seem to have been neglected from the very beginning in the generalization to Anderson Localization \cite{chalker}
and to Many-Body-Localization \cite{serbyn}. Here we prefer to keep them for theoretical consistency,
since they are necessary to have a stationary distribution for the disorder variables $h_i$.

\subsection{ Reminder on the properties of a single Ornstein-Uhlenbeck process }

 The Langevin equation
\begin{eqnarray}
\frac{d h(t)}{dt} && = -\frac{h(t)}{\tau} + \xi(t)
\label{langevinanderson}
\end{eqnarray}
with the  white noise $\xi(t) $ defined by its generating functional
\begin{eqnarray}
< e^{i \int_{-\infty}^{+\infty} dt g(t) \xi(t) }>  = e^{ -  \frac{W^2}{\tau}\int_{-\infty}^{+\infty} dt g^2(t) } 
\label{geneBB}
\end{eqnarray}
with zero average and delta two-point-correlation
\begin{eqnarray}
< \xi(t)  > && = 0
\nonumber \\
< \xi(t) \xi(t')) > && = 2 \frac{W^2}{\tau}   \delta(t-t')
\label{correbb}
\end{eqnarray}
can be integrated to obtain
\begin{eqnarray}
 h(t)= \int_{-\infty}^t dt'  \xi(t') e^{- \frac{t-t'}{\tau } }
\label{langevininteg}
\end{eqnarray}
The average vanishes
\begin{eqnarray}
 <h(t)> && =0 
\label{moy0}
\end{eqnarray}
and the two-point correlation decays exponentially
\begin{eqnarray}
< h(t_1) h(t_2)  > && = \int_{-\infty}^{t_1} dt  e^{- \frac{t_1-t}{\tau } }
\int_{-\infty}^{t_2} dt'   e^{- \frac{t_2-t'}{\tau } }<  \xi(t) \xi(t') >
 =2  \frac{W^2}{\tau}  \int_{-\infty}^{min(t_1,t_2)} dt  e^{ \frac{2t-t_1-t_2}{\tau } }
\nonumber \\
&& = W^2  e^{- \frac{ \vert t_1-t_2 \vert}{\tau } }
\label{hcorre}
\end{eqnarray}
on the time scale $\tau$.
The full generating functional reads
\begin{eqnarray}
 < e^{i \int_{-\infty}^{+\infty} dt g(t) h(t) }>  =
 e^{ -  \frac{W^2}{2} \int_{-\infty}^{+\infty} dt_1
 \int_{-\infty}^{+\infty} dt_2 g(t_1) g(t_2)
 e^{- \frac{ \vert t_1-t_2 \vert}{\tau }}}
\label{geneOrnstein}
\end{eqnarray}

The corresponding Fokker-Planck equation for the probability $P_t(h)$
to have the value $h$ at time $t$ reads
\begin{eqnarray}
\partial_t P_t(h) && = -  \partial_{h} 
\left[ \left(- \frac{h}{\tau} \right) P_t(h)    \right] +  \frac{W^2}{\tau}\partial_{h}^2 P_t(h)
\nonumber \\
&& =\frac{1}{\tau}  \partial_{h} \left[ h P_t(h) + W^2 \partial_{h} P_t(h)  \right]
\label{fokkerplanck1var}
\end{eqnarray}
so that the stationary distribution
\begin{eqnarray}
 P_*(h) && = \frac{1}{\sqrt{2 \pi W^2} } e^{- \frac{h^2}{2 W^2}}
\label{stattiogauss}
\end{eqnarray}
coincides with the disorder distribution of Eq. \ref{gauss}.

\subsection{ Stochastic dynamics for the Hamiltonian in the fictitious time $t$ }

In summary, the random Hamiltonian of Eq. \ref{H} has been promoted to
the stochastic process in the fictitious time $t$
\begin{eqnarray}
H (t) = H_0 + \sum_{i=1}^N h_i(t) O_i
\label{Ht}
\end{eqnarray}
and evolves according to
\begin{eqnarray}
\frac{d H(t)}{dt} && = \sum_{i=1}^N \frac{d h_i(t)}{dt} O_i
 = - \frac{1}{\tau}  \sum_{i=1}^N h_i(t)  O_i + \sum_{i=1}^N O_i \xi_i(t)
\label{Htderi}
\end{eqnarray}
with the $N$ white noises $ \xi_i(t)$ of Eq. \ref{correbbxi}.
The only new parameter is the correlation time $\tau$ governing the speed of the dynamics within
the space of disorder realizations. Its choice as a function of the system size will be discussed later.

\subsection{ Perturbation theory for a small time interval $\Delta t$ } 

Let us introduce the spectral decomposition of the Hamiltonian $H(t)$ at time $t$
\begin{eqnarray}
H(t) = \sum_{n=1}^{M} E_n(t) \vert \phi_n(t)> < \phi_n(t) \vert
\label{tighttimedt}
\end{eqnarray}
where the $E_n(t) $ are the $M$ energy levels, with their associated normalized eigenvectors $\vert \phi_n(t)> $.

To compute the new eigenvalues $E_n(t+\Delta t)$ at time $t+\Delta t$
where the Hamiltonian reads
\begin{eqnarray}
H(t+\Delta t) 
&& =H(t)  + \Delta H(t)
\nonumber \\
  \Delta H(t) && =  -  \frac{\Delta t  }{\tau}  \sum_{i=1}^N h_i(t)  O_i
+ \sum_{i=1}^N O_i \int_t^{t+\Delta t} \xi_i(\tau) d\tau
\label{hnext}
\end{eqnarray}
one wishes to apply the standard perturbation theory to the difference  $\Delta H(t) $,
with the matrix elements
\begin{eqnarray}
&& < \phi_n(t) \vert \Delta H(t)  \vert \phi_m(t)>  
 =  -\frac{\Delta t  }{\tau}\sum_{i=1}^N h_i(t) < \phi_n(t)\vert  O_i \vert \phi_m(t)>  +
\sum_{i=1}^N < \phi_n(t)\vert O_i \vert \phi_m(t)>
 \int_t^{t+\Delta t}  \xi_i(\tau)  d\tau 
\label{hnextmat}
\end{eqnarray}
The first term is of order $\Delta t$, while the second term is of order $\sqrt{\Delta t}$,
so one uses the second order perturbation formula to obtain
\begin{eqnarray}
&& E_n(t+\Delta t)  = E_n(t) + < \phi_n(t) \vert \Delta H(t) \vert \phi_n(t)>
+ \sum_{m \ne n} \frac{< \phi_n(t) \vert \Delta H(t) \vert \phi_m(t)>
< \phi_m(t) \vert  \Delta H(t) \vert \phi_n(t)> }{E_n(t)- E_m(t)}
\nonumber \\
&& =  E_n(t) 
 - \frac{\Delta t  }{\tau} \sum_{i=1}^N h_i(t) < \phi_n(t)\vert  O_i \vert \phi_n(t)> 
 +
\sum_{i=1}^N < \phi_n(t)\vert O_i \vert \phi_n(t)>
 \int_t^{t+\Delta t}  \xi_i(\tau)  d\tau 
\nonumber \\
&&+ \sum_{m \ne n} \frac{
\left[ \sum_{i=1}^N < \phi_n(t)\vert O_i \vert \phi_m(t)>
 \int_t^{t+\Delta t}  \xi_i(\tau)  d\tau \right]
\left[ \sum_{j=1}^N < \phi_m(t)\vert O_j \vert \phi_n(t)>
 \int_t^{t+\Delta t}  \xi_j(\tau')  d\tau' \right] }
{E_n(t)- E_m(t)}
+o(\Delta t)
\label{ener2d}
\end{eqnarray}

\subsection{ Langevin Equations  for the energy levels }

In the limit $\Delta t \to 0 $, one obtains the Langevin equations
\begin{eqnarray}
\frac{d E_n(t)}{dt} && = F_n(t)  + \Lambda_n(t)
\label{langevinE}
\end{eqnarray}
where the forces $F_n(t) $ are computed from
the averages over the white noises $\xi_i(t)$
\begin{eqnarray}
 F_n(t) && \equiv  \lim \limits_{\Delta t \to 0}  \frac{<E_n(t+\Delta t)- E_n(t)>}{\Delta t} 
\nonumber \\
&& = -  \frac{1  }{\tau} \sum_{i=1}^N h_i(t) < \phi_n(t)\vert  O_i \vert \phi_n(t)> 
  + 2  \frac{W^2}{\tau}\sum_{m \ne n} 
 \frac{ \sum_{i=1}^N  \vert < \phi_n(t)\vert O_i \vert \phi_m(t)> \vert^2 }
{E_n(t)- E_m(t)}
\label{mun}
\end{eqnarray}
and where the Langevin noises read
\begin{eqnarray}
\Lambda_n(t) \equiv \sum_{i=1}^N  < \phi_n(t)\vert O_i \vert \phi_n(t)> \xi_i(t)  
\label{zetan}
\end{eqnarray}
so that their average values vanish
\begin{eqnarray}
< \Lambda_n(t) >
&& =0
\label{zetanave}
\end{eqnarray}
and their two-point correlations read
\begin{eqnarray}
< \Lambda_n(t)\Lambda_m(t') >
&& =\sum_{i=1}^N  < \phi_n(t)\vert O_i \vert \phi_n(t)> 
\sum_{j=1}^N  < \phi_m(t)\vert O_j \vert \phi_m(t)>
< \xi_i(t)\xi_j(t')>
\nonumber \\   
&& = 2  \delta(t-t') \frac{W^2}{\tau} \sum_{i=1}^N 
 < \phi_n(t)\vert O_i \vert \phi_n(t)> < \phi_m(t)\vert O_i \vert \phi_m(t)>
\label{zetancorre}
\end{eqnarray}

\subsection{ Rescaling the energies with the level spacing $\Delta_N $  }

It is convenient to rescale the energies 
\begin{eqnarray}
E_n(t)= \Delta_N e_n(t)
\label{erescal}
\end{eqnarray}
with the level spacing $\Delta_N $ in the middle of the spectrum,
in order to work with rescaled energies $e_n(t)$ of order $O(1)$.

Then the Langevin Eq. \ref{langevinE} become for the rescaled energies
$e_n(t)$ 
\begin{eqnarray}
 \frac{d e_n(t)}{dt} && =  f_n(t) +  \lambda_n(t)
\label{langevinerescal}
\end{eqnarray}
with the rescaled forces
\begin{eqnarray}
f_n(t) \equiv \frac{ F_n(t) }{\Delta_N }  
&& = - \frac{ 1  }{ \tau \Delta_N }
 \sum_{i=1}^N h_i(t) < \phi_n(t)\vert  O_i \vert \phi_n(t)> 
  +   \frac{ 2  }{\tau \Delta_N^2 } W^2 \sum_{m \ne n} 
 \frac{ \sum_{i=1}^N  \vert < \phi_n(t)\vert O_i \vert \phi_m(t)> \vert^2 }
{e_n(t)- e_m(t)}
\label{frescal}
\end{eqnarray}
and the rescaled Langevin noises
\begin{eqnarray}
 \lambda_n(t) && \equiv  \frac{ \Lambda_n(t) }{\Delta_N }  
\label{lambdarescal}
\end{eqnarray}
with the two-point correlations
\begin{eqnarray}
< \lambda_n(t)\lambda_m(t') > = \frac{ < \Lambda_n(t)\Lambda_m(t') >}{\Delta_N^2}
&&  = 2  \delta(t-t') \frac{ W^2}{\tau \Delta_N^2 }  \sum_{i=1}^N 
 < \phi_n(t)\vert O_i \vert \phi_n(t)> < \phi_m(t)\vert O_i \vert \phi_m(t)>
\label{correlambdarescal}
\end{eqnarray}

\section{ Application to Many-Body-Localization Hamiltonians }

\label{sec_mbl}

\subsection{ Quantum spin chain with random fields }

In this section, following \cite{serbyn},
 we consider the quantum spin chain of $N$
 sites with an Hilbert space of size $M=2^N$
of the form of Eq. \ref{H} with $O_i=\sigma_i^z$
\begin{eqnarray}
H = H_0 + \sum_{i=1}^N h_i \sigma_i^z
\label{Hmbl}
\end{eqnarray}
The non-random part can be for instance the Heisenberg Hamiltonian,
since this is the model where the MBL transition has been studied numerically on
the largest sizes \cite{alet}.

In the middle of the spectrum, the level spacing scales as
\begin{eqnarray}
 \Delta_N = N^{\frac{1}{2}} 2^{-N}
\label{deltan}
\end{eqnarray}

\subsection{ The doubly stochastic Edwards-Anderson matrix }

Let us now analyze the matrix elements appearing in the Langevin equations.

In the middle of the spectrum,
the matrix elements of the magnetization $\sigma_i^z$ at a given point $i$ 
\begin{eqnarray}
m_{nm} [i] &&  \equiv  < \phi_n(t)\vert \sigma_i^z \vert \phi_m(t)>
\label{mni}
\end{eqnarray}
are expected to be of random sign, uncorrelated with the sign of $h_i(t)$.
It is thus appropriate to consider their squares,
 i.e. the Edwards-Anderson order
parameters as in the field of spin-glasses \cite{EA}.

The local Edwards-Anderson matrix 
\begin{eqnarray}
q^{EA}_{nm}[i] && 
 \equiv   \vert < \phi_n(t)\vert \sigma_i^z \vert \phi_m(t)>\vert^2= < \phi_n(t)\vert \sigma_i^z \vert \phi_m(t)> 
< \phi_m(t)\vert \sigma_i^z \vert \phi_n(t)> 
\label{qeanmloci}
\end{eqnarray}
is doubly stochastic, i.e. it is a square matrix of size $M \times M$
of non-negative real numbers,
where the sums over any row or any column is unity
\begin{eqnarray}
\sum_{n=1}^M q^{EA}_{nm} [i] = 1 = \sum_{m=1}^M q^{EA}_{nm} [i]
\label{bistochloci}
\end{eqnarray}
as a consequence of the completeness identity
\begin{eqnarray}
\sum_{n=1}^M \vert \phi_n(t)> < \phi_n(t)\vert = Id
\label{completude}
\end{eqnarray}
and the Pauli matrix identity $(\sigma_i^z)^2=1$.
Doubly stochastic matrices appear very often in quantum mechanics 
(see for instance \cite{rigol,louck,luck}).

The global Edwards-Anderson matrix of elements
\begin{eqnarray}
q^{EA}_{nm} (N) &&  \equiv \frac{ 1}{ N }  \sum_{i=1}^N \vert < \phi_n(t)\vert \sigma_i^z \vert \phi_m(t)>\vert^2 = \frac{ 1}{ N }  \sum_{i=1}^N q^{EA}_{nm}[i]
\label{qeanmbis}
\end{eqnarray}
is thus also doubly stochastic
\begin{eqnarray}
\sum_{n=1}^M q^{EA}_{nm} [N] = 1 = \sum_{m=1}^M q^{EA}_{nm} [N]
\label{bistoch}
\end{eqnarray}

In the thermodynamic limit $N \to +\infty$, 
the spatial average of Eq. \ref{qeanmbis} 
can be considered as equivalent to a disorder-average for the local Edwards Anderson matrix element
\begin{eqnarray}
q^{EA}_{nm} (N) &&  \simeq \overline{ q^{EA}_{nm}[i] }
\label{qeanmbisav}
\end{eqnarray}
In the middle of the spectrum, one expects that the diagonal Edwards-Anderson spin-glass order parameter 
within an eigenstate $n$  will not depend on the precise eigenstate $n$
\begin{eqnarray}
q^{EA}_{nn} (N) && \equiv \frac{ 1}{ N }  \sum_{i=1}^N \vert < \phi_n(t)\vert \sigma_i^z \vert \phi_n(t)> \vert^2 \simeq 
q^{EA}_{diag} (N)
\label{qdiag}
\end{eqnarray}
When the two states differ $n \ne m$, we keep the notation $q^{EA}_{nm} (N) $
and postpone the discussion on the dependence with respect to the positions of the two eigenstates.

\subsection{ Matrix elements involved in the Langevin noises }

The diagonal Edwards-Anderson spin-glass order parameter of Eq. \ref{qdiag}
directly appears in the diagonal correlations of the Langevin noises 
of Eq. \ref{correlambdarescal}
\begin{eqnarray}
< \lambda_n(t)\lambda_n(t') > 
&&  = 2  \delta(t-t') \frac{ W^2}{\tau \Delta_N^2 }  N q^{EA}_{diag} (N) 
\label{correlambdarescalq}
\end{eqnarray}
The off-diagonal correlations of Eq. \ref{correlambdarescal}
for $n \ne m$
 involve a sum over the sites $i=1,2,..,N$ 
of terms $m_{nn} (i) m_{mm} (i) $ that are of random signs 
 in the middle of the spectrum
(For the ergodic phase, where E.T.H. at infinite temperature means that eigenvectors are like random vectors in the Hilbert space, 
the explicit computation of the vanishing 
correlation in the thermodynamic limit between 
the magnetizations $m_{nn} (i) $ and $ m_{mm} (i) $ on the same site
for two different eigenvectors $n \ne m$ is given in Appendix A section 3).
So the order of magnitude of this sum of terms $m_{nn} (i) m_{mm} (i) $
 of zero mean may be evaluated from its variance 
\begin{eqnarray}
 \sum_{i=1}^N m_{nn} (i) m_{mm} (i) \simeq \pm \sqrt{\sum_{i=1}^N m_{nn}^2 (i) m_{mm}^2 (i)  }
\simeq \pm \sqrt{ N  q^{EA}_{nn} (N) q^{EA}_{mm} (N) } \simeq
 \pm \sqrt{ N }  q^{EA}_{diag} (N) 
\label{sumoff}
\end{eqnarray}
Thus the off-diagonal correlation for $n \ne m$
\begin{eqnarray}
< \lambda_n(t)\lambda_m(t') > 
&&  =  \pm 2  \delta(t-t') \frac{ W^2}{\tau \Delta_N^2 } \sqrt{ N }  q^{EA}_{diag} (N)  
\label{correlambdarescalnegli}
\end{eqnarray}
are suppressed by a factor $1/\sqrt{ N }$ 
with respect to the diagonal correlations of Eq. \ref{correlambdarescalq}.
In the following, the off-diagonal correlations are thus neglected,
i.e. Eq. \ref{correlambdarescal} is replaced by
\begin{eqnarray}
< \lambda_n(t)\lambda_m(t') >  \opsimeq_{N \to +\infty}
&&  = 2  \delta(t-t') \delta_{n,m} \frac{ W^2}{\tau \Delta_N^2 }
  N q^{EA}_{diag} (N)
\label{correlambdarescaldiag}
\end{eqnarray}

\subsection{ Matrix elements involved in the Langevin forces } 

The second term of the forces of Eq. \ref{frescal}
involves the off-diagonal value $q^{EA}_{n \ne m} (N)$.

 The order of magnitude of the first term in Eq. \ref{frescal}
corresponding to a sum of terms of random signs 
may be evaluated by 
\begin{eqnarray}
 \sum_{i=1}^N h_i(t) m_{nn} (i) \simeq \pm \sqrt{\sum_{i=1}^N h^2_i(t) m_{nn}^2 (i)} 
\simeq \pm \sqrt{ W^2 \sum_{i=1}^N  m_{nn}^2 (i)} \simeq \pm \sqrt{ W^2 N q^{EA}_{nn} (N) } 
\simeq \pm \sqrt{ W^2 N q^{EA}_{diag} (N) } 
\label{frescal1}
\end{eqnarray}

\subsection { Choice of the correlation time $\tau_N$ as a function of the system-size $N$ }

The off-diagonal Edwards-Anderson cannot be greater than the diagonal order parameter of Eq. \ref{qdiag}
\begin{eqnarray}
q^{EA}_{n \ne m} (N)   \leq   q^{EA}_{diag} (N)
\label{offsmaller}
\end{eqnarray}
As a consequence, to obtain Langevin equations independent of the system size $N$ in the thermodynamic limit $N \to +\infty$ in the middle of the spectrum, the correlation time $\tau_N$ has to be chosen to make the amplitude of the noise in Eq. \ref{correlambdarescaldiag} size-independent with some fixed $\tau_0$
\begin{eqnarray}
< \lambda_n(t)\lambda_m(t') >  \opsimeq_{N \to +\infty}
&&  = 2  \delta(t-t') \delta_{n,m} \frac{ W^2}{\tau_0  }  
\label{correlambdarescaldiag0}
\end{eqnarray}
Using Eq. \ref{deltan} for the level spacing $\Delta_N$, one thus obtains
the choice
\begin{eqnarray}
\tau_N = \tau_0 \frac{N q^{EA}_{diag} (N) }{ \Delta_N^2 }
  = \tau_0 2^{2N} q^{EA}_{diag} (N) 
\label{choicetau}
\end{eqnarray}
The physical meaning of this scaling is related to the adiabatic theorem :
the eigenstates of the time-varying Hamiltonian $H(t)$ can be followed
in time
only if the dynamics is sufficiently slow with respect to the spectrum.

With this choice, the force of Eq. \ref{frescal} reads
\begin{eqnarray}
f_n(t) && = - \frac{ 1  }{ \tau_N \Delta_N }
 \sum_{i=1}^N h_i(t)  m_{nn} (i) 
  +   \frac{ 2  }{\tau_N \Delta_N^2 } W^2 \sum_{m \ne n} 
 \frac{  N q^{EA}_{nm} (N) }
{e_n(t)- e_m(t)}
\nonumber \\ 
&& =- \frac{ 1 }{ \tau_0 N^{\frac{1}{2}}  2^{N} q^{EA}_{diag} (N) }
 \sum_{i=1}^N h_i(t)  m_{nn} (i) 
  +   \frac{ 2  }{\tau_0  q^{EA}_{diag} (N) } W^2 \sum_{m \ne n} 
 \frac{   q^{EA}_{nm} (N) }
{e_n(t)- e_m(t)}
\label{frescaltau}
\end{eqnarray}
From the estimation of the amplitude of Eq. \ref{frescal1},
one obtains that the order of magnitude of the first term is
\begin{eqnarray}
\frac{ 1 }{ \tau_0 N^{\frac{1}{2}}  2^{N} q^{EA}_{diag} (N) } \sum_{i=1}^N h_i(t) m_{nn} (i) \simeq \pm 
\frac{ W }{ \tau_0  2^{N} \sqrt{   q^{EA}_{diag} (N) }    }
\label{frescal1bis}
\end{eqnarray}
Using Eq. \ref{bistoch} and Eq. \ref{offsmaller}, one obtains the bound
\begin{eqnarray}
 q^{EA}_{diag} (N) \geq \frac{1}{M} = 2^{-N}
\label{bound}
\end{eqnarray}
As a consequence, the denominator of Eq. \ref{frescal1bis} is always greater 
than $\sqrt{2^N}$, so the first term of the force is always negligible in the thermodynamic limit.

So in the thermodynamic limit, the Langevin forces read
\begin{eqnarray}
f_n(t) && =   \frac{ W^2 }{\tau_0  }  \sum_{m \ne n} 
 \frac{ \beta_{nm}(N)  } {e_n(t)- e_m(t)}
\label{ffinal}
\end{eqnarray}
where we have introduced the ratios 
\begin{eqnarray}
\beta_{nm}(N) = \equiv \frac{ 2 q^{EA}_{nm} (N) }{  q^{EA}_{diag} (N) } = \beta_{mn}(N)
\label{betan}
\end{eqnarray}

\subsection{ Fokker-Planck equation for the energy levels   }

Let us summarize the output of the previous sections.
The Langevin equations for the rescaled energies $e_n(t)$ read
\begin{eqnarray}
 \frac{d e_n(t)}{dt} && =  f_n(t) +  \lambda_n(t)
\label{langevinfinal}
\end{eqnarray}
with the Langevin noises $\lambda_n(t) $ of correlations
\begin{eqnarray}
< \lambda_n(t)\lambda_m(t') >    = 2  \delta(t-t') \delta_{n,m} \frac{ W^2}{\tau_0  }  
\label{fincorre}
\end{eqnarray}
provided the choice of the time correlation of Eq. \ref{choicetau} for the fictitious dynamics,
and the forces
\begin{eqnarray}
f_n(t) && =  \frac{ W^2  }{\tau_0   }  \sum_{m \ne n} \frac{  \beta_{nm}(N)  }{e_n(t)- e_m(t)}
\label{fffinal}
\end{eqnarray}
where the important parameters are the $\beta_{nm}(N)$ of Eq. \ref{betan}

The corresponding Fokker-Planck equation for the probability $P_t(e_1,..,e_M)$
to have the energies $(e_1,..,e_M)$ at time $t$ reads
\begin{eqnarray}
\partial_t P_t(e_1,..,e_M) && = - \sum_{n=1}^M \partial_{e_n} 
\left[f_n(e_1,..,e_M) P_t(e_1,..,e_M)    \right]
+ \frac{W^2}{ \tau_0} \sum_{n=1}^M    \partial_{e_n}^2  P_t(e_1,..,e_N)
\nonumber \\
 && =\frac{ W^2  }{\tau_0   } \sum_{n=1}^M \partial_{e_n} 
\left[  - \left( \sum_{m \ne n} \frac{  \beta_{nm}(N)  }{e_n(t)- e_m(t)} \right) P_t(e_1,..,e_M) 
+ \partial_{e_n}  P_t(e_1,..,e_N) \right]
\label{fokkerplanck}
\end{eqnarray}
so that the stationary distribution reads in the middle of the spectrum
\begin{eqnarray}
 P_*(e_1,..,e_M) && \propto
\prod_{n<m} \vert e_n-e_m \vert^{\beta_{nm}(N)} 
\label{statio}
\end{eqnarray}
i.e. the level repulsion between the two eigenvalues $e_n$ and $e_m$ 
is governed by the ratio $\beta_{nm}(N)$ of Eq. \ref{betan} between 
the corresponding Edwards-Anderson matrix elements.

\subsection{ Analysis in the delocalized phase }

In the delocalized phase where the Eigenstate Thermalization Hypothesis
(E.T.H.) holds, eigenstates in the middle of the spectrum
can be approximated by random vectors in the Hilbert space of size $M=2^N$.

Then one obtains (see Appendix A) 
that the the diagonal value $q_{diag}^{deloc}(N)$
(Eq \ref{qdiagdelocapp}) 
and the off-diagonal value (Eq \ref{qoffdelocapp})
read
\begin{eqnarray}
q_{diag}^{EA}(N) && \opsimeq_{N \to +\infty}   \frac{2}{2^N}
\nonumber \\
q^{EA}_{n \ne m} (N)  &&  \opsimeq_{N \to +\infty}  \frac{1}{2^N}
\label{qdiagdeloc}
\end{eqnarray}
i.e. they both decay as $1/2^N$, and the ratio of Eq. \ref{betan}
takes the simple GOE value as it should
\begin{eqnarray}
\beta_{nm}^{deloc} (N) = \frac{ 2 q^{EA}_{n \ne m} (N) }{  q^{EA}_{diag} (N) }\oppropto_{N \to +\infty}  1 
\label{betandeloc}
\end{eqnarray}

In addition, the behavior of $q_{diag}^{EA}(N) \propto 2^{-N} $ yields that
 the correlation time of Eq. \ref{choicetau} scales as
\begin{eqnarray}
\tau_N^{deloc}    = \tau_0 2^{2N} q^{EA}_{diag} (N) = 2 \tau_0 2^{N}
\label{choicetaudeloc}
\end{eqnarray}

\subsection{  Analysis in the localized phase      }

Let us first consider the infinitely strong localized phase, 
where the $2^N$ eigenstates are simply given by tensor products
in the $\sigma^z$ basis
\begin{eqnarray}
 \vert \phi_n(t)>  && =   \vert S_1,..,S_N >
\label{philoc}
\end{eqnarray}
Then the diagonal Edwards-Anderson order parameter
 reaches its maximal value unity
\begin{eqnarray}
q^{EA}_{diag} (N) &&  \equiv \frac{ 1}{ N }  \sum_{i=1}^N \vert < \phi_n(t)\vert \sigma_i^z \vert \phi_n(t)>\vert^2 = 1
\label{qeannloc}
\end{eqnarray}
whereas the off-diagonal Edwards-Anderson order parameters
completely vanish
\begin{eqnarray}
q^{EA}_{n \ne m} (N) &&  \equiv \frac{ 1}{ N }  \sum_{i=1}^N \vert < \phi_n(t)\vert \sigma_i^z \vert \phi_m(t)>\vert^2 = 0
\label{qeanmloc}
\end{eqnarray}
so the ratio of Eq. \ref{betan} also vanishes and gives the Poisson value
with no level repulsion as it should
\begin{eqnarray}
\beta_{nm}^{strongloc}(N) =0 
\label{betanloc}
\end{eqnarray}

Here the correlation time of Eq. \ref{choicetau}
scales as
\begin{eqnarray}
\tau_N^{strong loc} = \tau_0 2^{2N} 
\label{choicetaustrongloc}
\end{eqnarray}
i.e. it is much bigger than in the delocalized phase (Eq. \ref{choicetaudeloc}).

In the localized phase with an arbitrary finite localization length $\xi$,
the simplest guess is that each region of length $\xi$ with an Hilbert space of size $2^{\xi}$ is 'delocalized' leading to a diagonal Edwards-Anderson order parameter of order
(Eq. \ref{qdiagdeloc} with the replacement $N \to \xi$)
\begin{eqnarray}
q_{diag}^{EA}(N) && \simeq   \frac{2}{2^{\xi}}
\label{qdiagxi}
\end{eqnarray}
Then the double stochasticity condition of Eq. \ref{bistoch}
 leads to the 
following generic exponential decay for the off-diagonal value
\begin{eqnarray}
 q^{EA}_{n \ne m} (N) \propto \frac{1 }{2^N}
\label{qoffxi}
\end{eqnarray}
Another way to arrive at the same conclusion is explained in Ref \cite{papic} 
in terms of the Local Integrals of Motion (LIOMs) that characterize the Many-Body Localized phase :
two generic eigenstates have different LIOMs everywhere, so the matrix element of a local operator
involves the tunneling of the excitation through the entire system, and this tunneling is exponentially suppressed.

The exponentially rare pairs of eigenstates $(n,m)$ that can have a finite off-diagonal value $q^{EA}_{n \ne m} (N) $
are the states corresponding to only one or a few different LIOMS in the region of the local operator,
but these states are not consecutive levels in the spectrum of the whole system.

Eq \ref{qoffxi} yields
  that the level repulsion index
  converges exponentially towards zero (Eq. \ref{betanloc}) as
\begin{eqnarray}
 \beta_{nm}^{loc} (N)\propto  2^{-(N-\xi) }
\label{betaxi}
\end{eqnarray}
i.e the finite-size scaling would not involve the 
standard ratio of the lengths $N/\xi$,
but the ratio of the sizes of the corresponding 
Hilbert spaces $2^N/2^{\xi} $.

\subsection{  Analysis at the critical point    }

At criticality where the localization length $\xi$ diverges, one expects that the diagonal value
$q_{diag}^{EA}(N)$ does not remain finite in the thermodynamic limit $N \to + \infty$,
and that the off-diagonal value $q_{n,n+1}^{EA}(N)$ between two consecutive levels
diplays the same size-decay as the diagonal value. Their ratio then produces  a non-trivial value
for the level repulsion exponent
\begin{eqnarray}
0< \beta_c = \frac{ 2 q^{EA}_{n,n+1} (N) }{  q^{EA}_{diag} (N) } <1
\label{betacr}
\end{eqnarray}
that can be anywhere between the Poisson limit $\beta=0$ and the
Wigner-Dyson limit $\beta=1$ depending on the MBL model.

In analogy with the Anderson Localization Transition case described in section \ref{anderson_criti},
it is tempting to speculate as in Ref \cite{serbyn} that some multifractality appears in matrix elements.
Indeed, besides Anderson localization transitions
where multifractality has been much studied \cite{mirlinrevue},
multifractal properties are actually generic at random
critical points
\cite{Ludwig,Jac_Car,Ols_You,Cha_Ber,Cha_Ber_Sh,PCBI,BCrevue,Sourlas,Thi_Hil,Par_Sou,DPmultif,wettingmultif,nancy}. 
In Ref \cite{serbyn}, this multifractality is used to obtain that the critical off-diagonal value $q_{n,m}^{EA}(N)$ decays for large energy separation $\vert e_n -e_m \vert$, so that the effective repulsive interaction between eigenvalues
 becomes short-ranged with respect to the energy separation, which is necessary to obtain a non-vanishing compressibility $0<\chi_c<1$ and the exponent $\gamma_c=1$ at large distance in the semi-Poisson probability distribution of the level-spacing (see more details and numerical results in \cite{serbyn} in relation with the plasma model \cite{plasma}).

\section{ Dyson Brownian approach for Anderson localization models }

\label{sec_anderson}

\subsection{ Anderson Localization models }

In this section, the Dyson approach is applied
 to the Anderson Localization 
model for a single particle in a volume of $N=L^d$ sites
\begin{eqnarray}
H=H_0 + \sum_{i=1}^N h_i \vert i > < i \vert 
\label{handerson}
\end{eqnarray}
where the Gaussian $h_i$ are the random on-site energies,
and where $H_0$ contains the hopping terms.
This corresponds to Eq. \ref{H} with the local operators
$O_i=\vert i > < i \vert  $. The original application of
the Dyson approach to Anderson Localization models was formulated
with continuum space \cite{chalker}.

The Hilbert space is of size $M=N=L^d$, and the level spacing 
in the middle of the spectrum is 
\begin{eqnarray}
\Delta_N= \frac{1}{N}=\frac{1}{L^d}
\label{levelanderson}
\end{eqnarray}

\subsection{  Density Correlation matrix}

Here in the Langevin equations,
 the underlying doubly stochastic matrix is the Density Correlation matrix
\begin{eqnarray}
Y_{n,m}(N) \equiv \sum_{i=1}^N   \vert < \phi_n(t)\vert i>\vert^2 
\vert < i \vert \phi_m(t)>^2 \vert
\label{andersonbi}
\end{eqnarray}
that satisfies
\begin{eqnarray}
\sum_{n=1}^M Y_{n,m}(N)  = 1 = \sum_{m=1}^M Y_{n,m}(N)
\label{bistochloc}
\end{eqnarray}
as a consequence of the completeness identity for the basis of eigenstates
(Eq. \ref{completude}) and for the spatial basis
\begin{eqnarray}
\sum_{i=1}^N \vert i> < i\vert = Id
\label{completudei}
\end{eqnarray}
As stressed in \cite{luck}, the matrices of the form of Eq. \ref{andersonbi}
have the additional property to be factorizable
 into the product of a matrix $R$
and its transpose $R^t$
\begin{eqnarray}
Y_{n,m}(N) \equiv (R R^t)_{n,m} 
\label{yfactor}
\end{eqnarray}
where the matrix $R$, with left index corresponding to eigenstates $n$
and with right index corresponding to spatial positions $i$
\begin{eqnarray}
R_{n,i} \equiv  \vert < \phi_n(t)\vert i>^2 \vert
\label{rmatrix}
\end{eqnarray}
is also doubly stochastic
\begin{eqnarray}
\sum_{n=1}^N R_{n,i}  = 1 = \sum_{i=1}^N  R_{n,i}
\label{bistochlocrmatrix}
\end{eqnarray}

The diagonal elements are the well known 
Inverse Participation Ratios \cite{mirlinrevue} of index $q=2$ for a single eigenstate $\phi_n(t)$ :
again in the middle of the spectrum, we will consider that it does not depend
on the precise eigenstate $n$
\begin{eqnarray}
Y_{diag}(N) = Y_{n,n}(N) =  \sum_{i=1}^N   \vert < \phi_n(t)\vert i> \vert^{4} 
\label{yq}
\end{eqnarray}
They appear in the Langevin noise correlations for $n=m$
\begin{eqnarray}
< \lambda_n(t)\lambda_n(t') >
&&  = 2  \delta(t-t') \frac{ W^2}{\tau \Delta_N^2 } Y_{diag}(N)
\label{correandersondiag}
\end{eqnarray}

The off-diagonal value that represents the density correlation between
 different eigenstates $n \ne m$ \cite{chalker}
\begin{eqnarray}
  Y_{n \ne m}(N)= \sum_{i=1}^N   \vert < \phi_n(t)\vert i> \vert^{2} \vert < \phi_m(t)\vert i> \vert^{2} = Y_{m n}(N)
\label{zq}
\end{eqnarray}
appear both in the forces 
\begin{eqnarray}
f_n(t) && = - \frac{ 1  }{ \tau \Delta_N }
 \sum_{i=1}^N h_i(t) \vert < \phi_n(t)\vert i> \vert^2
  +   \frac{ 2  }{\tau \Delta_N^2 } W^2 \sum_{m \ne n} 
 \frac{  Y_{nm}(N) }
{e_n(t)- e_m(t)}
\label{fandersonz}
\end{eqnarray}
and in the off-diagonal correlations of the noise
\begin{eqnarray}
< \lambda_n(t)\lambda_m(t') >
&&  = 2  \delta(t-t') \frac{ W^2}{\tau \Delta_N^2 } Y_{nm}(N) 
\label{correandersonoffz}
\end{eqnarray}

\subsection { Choice of the correlation time $\tau_N$ as a function of the system-size $N$ }

The off-diagonal elements of Eq. \ref{zq} cannot be greater
 than the diagonal value $Y_{diag}$
\begin{eqnarray}
Y_{n \ne m} (N) \leq Y_{diag}(N)
\label{z2smaller}
\end{eqnarray}
As a consequence, to obtain Langevin equations independent of the system size $N$ in the thermodynamic limit $N \to +\infty$ in the middle of the spectrum, the correlation time $\tau_N$ has to be chosen to make the diagonal correlation
 of the noise in Eq. \ref{correandersondiag} size-independent
\begin{eqnarray}
< \lambda_n(t)\lambda_n(t') >  \opsimeq_{N \to +\infty}
&&  = 2  \delta(t-t')  \frac{ W^2}{\tau_0  }  
\label{correlambdarescaldiaga}
\end{eqnarray}
Using Eq. \ref{levelanderson} for the level spacing $\Delta_N$, one
obtains the choice
\begin{eqnarray}
\tau_N = \tau_0 \frac{ Y_{diag} (N) }{ \Delta_N^2 } =  \tau_0 N^2 Y_{diag}(N)
\label{choicetauanderson}
\end{eqnarray}
As before, the physical meaning of this scaling is the adiabatic theorem.

With this choice, the off-diagonal correlator reads
\begin{eqnarray}
< \lambda_n(t)\lambda_m(t') >
&&  = 2  \delta(t-t') \frac{ W^2}{\tau_0  } \frac{Y_{nm}(N)}{Y_{diag}(N)} 
\label{correandersonoffzy}
\end{eqnarray}
and the forces become
\begin{eqnarray}
f_n(t) 
&& = - \frac{ 1  }{ N \tau_0 Y_{diag}(N) }
 \sum_{i=1}^N h_i(t) \vert < \phi_n(t)\vert i> \vert^2
  +   \frac{ 2   }{\tau_0 Y_{diag}(N)  } W^2 \sum_{m \ne n} 
 \frac{ Y_{nm}(N)  } {e_n(t)- e_m(t)}
\label{fandersonzz}
\end{eqnarray}

Since the $h_i$ are Gaussian variables of random sign,
the amplitude of the first term 
can be estimated as
\begin{eqnarray}
 \frac{ 1  }{ N \tau_0 Y_{diag}(N) }
 \sum_{i=1}^N h_i(t) \vert < \phi_n(t)\vert i> \vert^2
  \simeq \pm   \frac{ 1  }{ N \tau_0 Y_{diag}(N) } \sqrt{ W^2 \sum_{i=1}^N  \vert < \phi_n(t)\vert i> \vert^4 }
= \pm \frac{ W  }{ N \tau_0 (Y_{diag}(N))^{\frac{1}{2}} } 
\label{fandersonz1}
\end{eqnarray}
Eq. \ref{bistochloc} and \ref{z2smaller} yields the bound
\begin{eqnarray}
  Y_{diag}(N) \geq \frac{1}{N}  
\label{boundy2}
\end{eqnarray}
so the denominator of Eq. \ref{fandersonz1} cannot be smaller than $N^{1/2}$.
The first term of the force can thus be neglected in the thermodynamic limit
$N \to +\infty$,
i.e. the forces reduce to
\begin{eqnarray}
f_n(t) 
&& =   \frac{ 2   }{\tau_0 Y_{diag}(N)  } W^2 \sum_{m \ne n} 
 \frac{ Y_{nm}(N)  } {e_n(t)- e_m(t)}
\label{fandersonzfinal}
\end{eqnarray}

\subsection { Fokker-Planck equation }

The Fokker-Planck equation associated to the above Langevin dynamics reads
\begin{eqnarray}
&& \frac{\tau_0}{W^2} \partial_t P_t(e_1,..,e_M) 
  =\sum_{n=1}^M \partial_{e_n} 
\left[  - \left( \frac{2 }{Y_{diag}(N)} \sum_{m \ne n} \frac{  Y_{nm}(N)  }{e_n(t)- e_m(t)} \right) P_t(e_1,..,e_M) \right]
\nonumber \\ && +  \sum_{n=1}^M \partial^2_{e_n}  P_t(e_1,..,e_N) 
+    \sum_{n \ne m} \partial_{e_n} \partial_{e_m} 
\left(\frac{ Y_{nm}(N)}{Y_{diag}(N)} P_t(e_1,..,e_N) \right)P_t(e_1,..,e_N) 
  \\
&& \nonumber = \sum_{n=1}^M \partial_{e_n} \left[ 
 -  \left( \sum_{m \ne n} \frac{ \left(\frac{2 Y_{nm}(N) }{Y_{diag}(N)}\right)   }{e_n(t)- e_m(t)} \right) P_t(e_1,..,e_M) 
+  \partial_{e_n}  P_t(e_1,..,e_N) 
+  \sum_{m \ne n}  \partial_{e_m} \left(\frac{ Y_{nm}(N)}{Y_{diag}(N)} P_t(e_1,..,e_N) \right) \right]
\label{fokkerplanckanderson}
\end{eqnarray}

In the middle of the spectrum, we look for a stationary distribution with some exponents $\beta_{nm}(N)$
\begin{eqnarray}
 P_*(e_1,..,e_M) && \propto
\prod_{n<m} \vert e_n-e_m \vert^{\beta_{nm}(N)} 
\label{statiobeta}
\end{eqnarray}
This form satisfies the properties
\begin{eqnarray}
 \partial_{e_m} P_*(e_1,..,e_M) && = P_*(e_1,..,e_M)   \sum_{m' \ne m} \frac{  \beta_{mm'}(N)  }{e_{m}(t)- e_{m'}(t)} 
\label{deri1}
\end{eqnarray}
and 
\begin{eqnarray}
\sum_{m \ne n}  \partial_{e_m}  P_*(e_1,..,e_M) && = - \partial_{e_n} P_*(e_1,..,e_M)
\label{deri1sum}
\end{eqnarray}
As a consequence, the stationary Fokker-Planck Equation is satisfied by the form of Eq. \ref{statiobeta}
with the exponents
\begin{eqnarray}
\beta_{nm}(N) =  \frac{2 Y_{nm}(N)}{Y_{diag}(N)- Y_{nm}(N)} 
\label{betaanderson}
\end{eqnarray}

\subsection { Scaling in the localized phase }

In the localized phase, the Inverse Participation Ratio $Y_{diag}(N)$,
which represents an order parameter of the Localized Phase, remains finite
in the thermodynamic limit
\begin{eqnarray}
Y_{diag}(N) \opsimeq_{N \to +\infty} Y_{diag}(\infty) >0
\label{yqloc}
\end{eqnarray}
whereas its off-diagonal version $Y_{n \ne m}(N)$ is expected
to vanish exponentially beyond the localization length $\xi$
\begin{eqnarray}
Y_{n \ne m}(N) \oppropto_{ L \gg  \xi} e^{- \frac{L}{\xi}}
\label{zqloc}
\end{eqnarray}
The rare pairs of eigenstates $(n,m)$ that can have a finite off-diagonal value $Y_{n \ne m} (N) $
are the states whose localization centers are separated by a distance smaller than the localization length $\xi$,
but they are not consecutive levels in the spectrum of the whole sample.

Eq. \ref{zqloc} yields the level repulsion exponent
 converges exponentially towards to the Poisson value zero
\begin{eqnarray}
\beta^{loc}_{nm} \propto e^{- \frac{L}{\xi}}
\label{betaandersonpoisson}
\end{eqnarray}

In addition the correlation time of Eq. \ref{choicetauanderson}
has to be chosen as
\begin{eqnarray}
\tau_N^{loc}  =  \tau_0 N^2 Y_{diag}(+\infty) \propto N^2
\label{choicetauandersonloc}
\end{eqnarray}

\subsection { Scaling in the delocalized phase }

In the delocalized phase, the Inverse Participation Ratio $Y_{diag}(N)$
 and the density correlation $Y_{n \ne m}(N)$
both vanish as $1/N$ with the numerical prefactors
 (see Eqs \ref{z2mdeloc} and \ref{y2mdeloc})
\begin{eqnarray}
Y^{deloc}_{diag}(N) && \opsimeq_{N \to +\infty} \frac{3}{N}
\nonumber \\
Y^{deloc}_{n \ne m}(N) && \opsimeq_{N \to +\infty} \frac{1}{N}
\label{yqdeloc}
\end{eqnarray}
Eq. \ref{betaanderson} then leads to the expected GOE value
\begin{eqnarray}
\beta^{deloc}_{nm} =  1
\label{betaandersongoe}
\end{eqnarray}

Note that the correlation time of Eq. \ref{choicetauanderson}
has to be chosen as
\begin{eqnarray}
\tau_N^{deloc}  =  \tau_0 N^2 Y_{diag}(N) \propto N
\label{choicetauandersondeloc}
\end{eqnarray}

\subsection { Scaling at criticality  }

\label{anderson_criti}

At criticality, the Inverse Participation Ratio
 $Y_{diag}(N)$ and the density correlation
$Y_{n,n+1}(N)$ for two consecutive levels are governed by the same exponent $\tau_2$ 
belonging to the multifractal spectrum $\tau_q$ defined for
the continuous index $q$ (see the review \cite{mirlinrevue} and references therein)
\begin{eqnarray}
Y_{diag}(N) \oppropto L^{-\tau_2}
\nonumber \\
 Y_{n,n+1}(N) \oppropto L^{-\tau_2}
\label{multif}
\end{eqnarray}

Their ratio leads to an intermediate non trivial value
for the level repulsion exponent
\begin{eqnarray}
0< \beta^c = \frac{2 Y_{n,n+1}(N)}{Y_{diag}(N)- Y_{n,n+1}(N)}  <1
\label{betacrloc}
\end{eqnarray}
that reflects the strength of the multifractality
of the wave-functions \cite{mirlinrevue}.
In the strong multifractality regime
where eigenfunctions are dominated by a few spikes, 
the off-diagonal value $Y_{n,n+1}(N)$ will
 remain small with respect to the diagonal
value $Y_{diag}(N) $ and one recovers the Poisson limit $\beta^c_{n,n+1} \simeq 0$.
On the contrary in the weak multifractality regime 
where eigenfunctions are nearly homogeneous, one recovers the Wigner Dyson limit $\beta^c_{n,n+1} \simeq 1$.

The behavior of the off-diagonal density correlation for larger energy spacing $\vert e_n -e_m \vert \geq 1$
(in terms of the rescaled energies $(e_n)$, the mean level spacing is unity)
follows the Chalker's scaling (see the review \cite{mirlinrevue} and references therein)
\begin{eqnarray}
 Y_{n,m}(N) \oppropto \frac{ L^{-\tau_2} }{ \vert e_n -e_m \vert^{1-\frac{\tau_2}{d} }}
\label{chalkerscaling}
\end{eqnarray}
This explains why there is no long-ranged interaction
between the eigenvalues at criticality
so that the critical point is characterized by
a non-vanishing compressibility $0<\chi_c<1$ and a semi-Poisson probability distribution of the level-spacing
with an exponential decay at large spacing.

\section{ Conclusion }

\label{sec_conclusion}

In summary, we have revisited the application of the Dyson Brownian Motion approach of random matrices \cite{dyson} to Anderson Localization (AL) models \cite{chalker} and to Many-Body Localization (MBL) Hamiltonians \cite{serbyn} in order to extract the level repulsion exponent $\beta$ in terms of the matrix elements
that appear in the Langevin and in the Fokker-Planck equations for the energy levels.
 For the MBL quantum spin Hamiltonian with random fields, we have obtained $\beta$
 in terms of the diagonal and off-diagonal matrix elements of the doubly stochastic Edwards-Anderson matrix.  For the Anderson Localization tight-binding models with random on-site energies, we have obtained $\beta $ in terms of the diagonal and off-diagonal matrix elements of the doubly stochastic Density Correlation matrix.

A very interesting issue is whether it could be possible to obtain from the Dyson approach other properties of the level statistics beyond the level repulsion index $\beta$,
like for instance the spectral compressibility $\chi$. Indeed at Anderson Localization Transitions,
the spectral compressibility $\chi$ has been first conjectured to be related to the correlation dimension $D_2$ of eigenstates
\cite{conjectured2} and later to the information dimension $D_1$ governing the entropy of eigenstates \cite{olivier_conjecture}.

For the MBL transitions, the Dyson Brownian Motion approach \cite{serbyn}
discussed here, and other arguments \cite{papic}
suggest that the key observables are the matrix elements of
local operators, so it would be very useful to better understand their behaviors and to measure them numerically besides the other interesting observables.

\appendix

\section{ Edwards-Anderson matrix for random eigenvectors }

\subsection{ Diagonal Edwards-Anderson order parameter for a single random vector }

Let us consider a random vector in the Hilbert space of size $M=2^N$
\begin{eqnarray}
 \vert \phi_n>  && =  \sum_{S_1,..,S_N} c_n(S_1,..,S_N) \vert S_1,..,S_N >
\label{randomsigndeloc}
\end{eqnarray}
where the only constraint is the normalization
\begin{eqnarray}
 1= \sum_{S_1,..,S_N} c_n^2(S_1,..,S_N) 
\label{randomsigndelocnorm}
\end{eqnarray}

We are interested into the magnetization $\sigma_i^z$ at some site $i$,
for instance $i=1$
\begin{eqnarray}
m && \equiv m_{nn}[i=1] = < \phi_n\vert \sigma_1^z \vert \phi_n>  = \sum_{S_1,..,S_N} S_1 c_n^2 (S_1,..,S_N) 
\nonumber \\&& = \sum_{S_2,..,S_N}  c_n^2 (S_1=+1,S_2..,S_N) - \sum_{S_2,..,S_N}  c_n^2 (S_1=-1,S_2..,S_N)
\label{m}
\end{eqnarray}

It is thus convenient to use the identity
\begin{eqnarray}
 1=\int_0^{+\infty} 2 R_n dR_n \delta \left(R_n^2- \sum_{S_2,..,S_N} c_n^2(S_1=+1,..,S_N) \right) 
\label{norma1}
\end{eqnarray}
to compute the distribution of $m$  up to some normalization 
\begin{eqnarray}
&& {\cal P}(m)  \propto \left[ \prod_{S_1,..,S_N} \int_{-\infty}^{+\infty} dc_n(S_1,..,S_N)
 \right] \delta\left(1-\sum_{S_1,..,S_N} c_n^2(S_1,..,S_N) \right) 
\int_0^{+\infty} 2 R_n dR_n \delta \left(R_n^2- \frac{1+m}{2} \right) 
\nonumber \\&&
\delta\left(m- \left(\sum_{S_2,..,S_N}  c_n^2 (S_1=+1,S_2..,S_N) - \sum_{S_2,..,S_N}  c_n^2 (S_1=-1,S_2..,S_N)\right)  \right)
\nonumber \\&&
\propto \int_0^{+\infty} 2 R_n dR_n \delta \left(R_n^2- \frac{1+m}{2} \right) 
\left[ \prod_{S_2,..,S_N} \int_{-\infty}^{+\infty} dc_n(S_1=+1,..,S_N)
\delta\left( R_n^2-\sum_{S_2,..,S_N}  c_n^2 (S_1=+1,S_2..,S_N)  \right) \right]
\nonumber \\&&
\times \left[ \prod_{S_2,..,S_N} \int_{-\infty}^{+\infty} dc_n(S_1=-1,..,S_N)
\delta\left( 1-R_n^2-\sum_{S_2,..,S_N} c_n^2(S_1=-1,S_2,..,S_N) \right) \right]
\label{pm}
\end{eqnarray}
The change of variables from the 
  $\frac{M}{2}=2^{N-1}$ Cartesian coordinates $c_n(S_1=+1,S_2,..,S_N) $ to the spherical coordinates of radius $\rho_n$ yields
\begin{eqnarray}
&&  \prod_{S_2,..,S_N} \int_{-\infty}^{+\infty} dc_n(S_1=+1,..,S_N)
\delta\left( R_n^2-\sum_{S_2,..,S_N} c_n^2(S_1=+1,S_2,..,S_N) \right) 
\nonumber \\
&& \propto \int_0^{+\infty} \rho_n^{\frac{M}{2}-1} d\rho_n
\delta\left(  R_n^2 -\rho_n^2 \right) 
 \propto \int_0^{+\infty}  \rho_n^{\frac{M}{2}-1} d\rho_n
 \frac{\delta\left(  R_n- \rho_n  \right) }{2 R_n}
 \propto  R_n^{\frac{M}{2}-2} 
\label{pmsphe}
\end{eqnarray}
Similarly
\begin{eqnarray}
 \prod_{S_2,..,S_N} \int_{-\infty}^{+\infty} dc_n(S_1=-1,..,S_N)
\delta\left( 1-R_n^2-\sum_{S_2,..,S_N} c_n^2(S_1=-1,S_2,..,S_N) \right)
 \propto  (\sqrt{1-R_n^2})^{\frac{M}{2}-2} 
\label{pmsphebis}
\end{eqnarray}

So the probability distribution of Eq. \ref{pm} reads for $-1 \leq m \leq 1$
\begin{eqnarray}
{\cal P}(m) 
&& \propto \int_0^{+\infty} 2 R_n dR_n \delta \left(R_n^2- \frac{1+m}{2} \right) 
R_n^{\frac{M}{2}-2}  (\sqrt{1-R_n^2})^{\frac{M}{2}-2} 
\nonumber \\
&& 
= \frac{ \Gamma \left(\frac{M}{4} +\frac{1}{2} \right)  }{{\sqrt \pi} \Gamma \left(\frac{M}{4}\right) }
 \left(1-m^2\right)^{\frac{M}{4}-1}  
\label{pp}
\end{eqnarray}
The average of course vanishes by symmetry, and its variance reads with $M=2^N$
\begin{eqnarray}
\overline{ m^2 } && = \frac{2}{2+M} =  \frac{2}{2+2^N}
\label{qdiagdelocm2}
\end{eqnarray}

In the delocalized phase, one thus expect that the diagonal Edwards-Anderson order parameter decays as
\begin{eqnarray}
q_{diag}^{deloc}(N) \oppropto_{N \to +\infty}    \frac{2}{2^N}
\label{qdiagdelocapp}
\end{eqnarray}

\subsection{ Off-diagonal Edwards-Anderson order parameter for two random orthogonal eigenvectors }

Here we consider two random orthogonal eigenvectors,
i.e. satisfying the normalization conditions
\begin{eqnarray}
 1= \sum_{S_1,..,S_N} c_n^2(S_1,..,S_N)  = \sum_{S_1,..,S_N} c_m^2(S_1,..,S_N) 
\label{norma2}
\end{eqnarray}
and the orthogonality condition
\begin{eqnarray}
0= \sum_{S_1,..,S_N}c_m(S_1,..,S_N) c_n(S_1,..,S_N) 
\label{orthog}
\end{eqnarray}

We are interested into the off diagonal matrix elements
of the magnetization on the site $i=1$
\begin{eqnarray}
&& v  \equiv   m_{mn}[i=1] = < \phi_m\vert \sigma_1^z \vert \phi_n>  = 
\sum_{S_1,..,S_N} S_1 c_m (S_1,..,S_N)  c_n (S_1,..,S_N)
\nonumber \\
&& = \sum_{S_2,..,S_N} c_m (S_1=+1,S_2..,S_N)  c_n (S_1=+1,S_2..,S_N)
 - \sum_{S_2,..,S_N}  c_m (S_1=-1,S_2..,S_N) c_n (S_1=-1,S_2..,S_N)
\label{elmn}
\end{eqnarray}

It is thus convenient to use the identities
\begin{eqnarray}
 1=\int_0^{+\infty} 2 R_n dR_n \delta \left(R_n^2- \sum_{S_2,..,S_N} c_n^2(S_1=+1,..,S_N) \right) 
\nonumber \\
 1=\int_0^{+\infty} 2 R_m dR_m \delta \left(R_m^2- \sum_{S_2,..,S_N} c_m^2(S_1=+1,..,S_N)  \right) 
\label{norma}
\end{eqnarray}
to compute the probability measure of $v$, up to some normalization 
\begin{eqnarray}
&& {\cal P}(v)  \propto 
\left[ \prod_{S_1,..,S_N}  \int_{-\infty}^{+\infty} dc_n(S_1,..,S_N)\int_{-\infty}^{+\infty} dc_m(S_1,..,S_N) \right]
\nonumber \\ 
&& \delta\left( 1- \sum_{S_1,..,S_N} c_n^2(S_1,..,S_N)  \right)
\delta \left( 1-  \sum_{S_1,..,S_N} c_m^2(S_1,..,S_N)   \right)
 \delta \left(  \sum_{S_1,..,S_N}c_m(S_1,..,S_N) c_n(S_1,..,S_N) \right )
 \nonumber \\ 
&&\delta \left( v- \left( \sum_{S_2,..,S_N} c_m (S_1=+1,..,S_N)  c_n (S_1=+1,..,S_N)
 - \sum_{S_2,..,S_N}  c_m (S_1=-1,..,S_N) c_n (S_1=-1,..,S_N) \right) \right )
 \nonumber \\ 
&&\propto \int_0^{+\infty} 2 R_n dR_n \int_0^{+\infty} 2 R_m dR_m
 \prod_{S_1,..,S_N} \left[ \int_{-\infty}^{+\infty} dc_n(S_1,..,S_N)\int_{-\infty}^{+\infty} dc_m(S_1,..,S_N) \right]
\nonumber \\ 
&& \delta \left(R_n^2- \sum_{S_2,..,S_N} c_n^2(S_1=+1,..,S_N) \right) 
\delta\left( 1-R_n^2- \sum_{S_2,..,S_N} c_n^2(S_1=-1,..,S_N)  \right)
\nonumber \\ 
&& \delta \left(R_m^2- \sum_{S_2,..,S_N} c_m^2(S_1=+1,..,S_N)  \right) 
\delta\left( 1-R_m^2- \sum_{S_2,..,S_N} c_m^2(S_1=-1,..,S_N)  \right)
 \nonumber \\ 
&&\delta \left( \frac{v}{2}-  \sum_{S_2,..,S_N} c_m (S_1=+1,S_2..,S_N)  c_n (S_1=+1,S_2..,S_N)  \right)
 \nonumber \\ 
&& \delta \left( \frac{v}{2}+\sum_{S_2,..,S_N}  c_m (S_1=-1,S_2..,S_N) c_n (S_1=-1,S_2..,S_N)  \right )
 \nonumber \\ 
&&\propto \int_0^{+\infty} 2 R_n dR_n \int_0^{+\infty} 2 R_m dR_m
G(R_n,R_m,v)  G(\sqrt{1-R_n^2 },\sqrt{1-R_n^2 },-v) 
\label{pv}
\end{eqnarray}
with the auxiliary function
\begin{eqnarray}
G(R_n,R_m,v) &&\equiv 
 \left[ \prod_{S_2,..,S_N}  \int_{-\infty}^{+\infty} dc_n(S_1=+1,..,S_N)\int_{-\infty}^{+\infty} dc_m(S_1=+1,..,S_N) \right]
\nonumber \\ 
&& \delta \left(R_n^2- \sum_{S_2,..,S_N} c_n^2(S_1=+1,..,S_N) \right) 
\nonumber \\ 
&& \delta \left(R_m^2- \sum_{S_2,..,S_N} c_m^2(S_1=+1,..,S_N)  \right) 
 \nonumber \\ 
&&\delta \left( \frac{v}{2}-  \sum_{S_2,..,S_N} c_m (S_1=+1,S_2..,S_N)  c_n (S_1=+1,S_2..,S_N)  \right)
\label{Gv}
\end{eqnarray}

For the $2^{N-1}=\frac{M}{2}$ Cartesian coordinates $c_n(S_1=+1,S_2..,S_N) $,
we go to the spherical coordinates of radius $\rho_n$.
For the $2^{N-1}=\frac{M}{2}$ Cartesian coordinates $c_m(S_1=+1,S_2..,S_N) $,
we go to the spherical coordinates of radius $\rho_m$,
but we need to keep the angle $\alpha$ of the scalar product
\begin{eqnarray}
\sum_{S_2,..,S_N} c_m (S_1=+1,S_2..,S_N)  c_n (S_1=+1,S_2..,S_N) = \rho_n \rho_m \cos \alpha
\label{theta}
\end{eqnarray}

So Eq. \ref{Gv} becomes up to angular normalization constants
\begin{eqnarray}
G(R_n,R_m,v) && \propto \int_0^{+\infty} \rho_{n}^{\frac{M}{2}-1} d\rho_{n} \delta \left(R_n^2- \rho_n^2 \right) 
\int_0^{+\infty} \rho_{m}^{\frac{M}{2}-1} d\rho_{m}  \delta \left(R_m^2-  \rho_{m}^2 \right) 
\int_0^{\pi} d \alpha (\sin \alpha)^{\frac{M}{2}-2}
\delta \left( \frac{v}{2}-  \rho_n \rho_m \cos \alpha  \right)
\nonumber \\ 
&&\propto \int_0^{+\infty} \rho_{n}^{\frac{M}{2}-1} d\rho_{n} 
\frac{ \delta \left(R_n- \rho_n \right) }{2 \rho_n }
\int_0^{+\infty} \rho_{m}^{\frac{M}{2}-1} d\rho_{m} \frac{ \delta \left(R_m- \rho_m \right) }{2 \rho_m } 
\int_0^{\pi} d \alpha (\sin \alpha)^{\frac{M}{2}-2}
\frac{\delta  \left(  \alpha- \arccos\frac{v}{2 R_n R_m }  \right)  }{ R_n R_m \sin \alpha }
\nonumber \\ 
&&\propto R_{n}^{\frac{M}{2}-3} 
R_{m}^{\frac{M}{2}-3}   
\left( \sqrt{1 - \frac{v^2}{4 R_n^2 R_m^2 }} \right)^{\frac{M}{2}-3}
 \theta \left(v^2 \leq 4 R_n^2 R_m^2 \right) 
\nonumber \\ 
&&\propto   
\left( \sqrt{ R_n^2 R_m^2 - \frac{v^2}{4 }} \right)^{\frac{M}{2}-3}
 \theta \left(v^2 \leq 4 R_n^2 R_m^2 \right) 
\label{Gvfin}
\end{eqnarray}

Eq. \ref{pv} then reads
\begin{eqnarray}
&& {\cal P}(v)  \propto \int_0^{1}  R_n dR_n \int_0^{1}  R_m dR_m
G(R_n,R_m,v)  G(\sqrt{1-R_n^2 },\sqrt{1-R_n^2 },-v) 
\nonumber \\ 
&& = {\cal N} \int_0^{1}  R_n dR_n \int_0^{1}  R_m dR_m
   \left( \sqrt{ R_n^2 R_m^2 - \frac{v^2}{4 }} \right)^{\frac{M}{2}-3}
 \theta \left(v^2 \leq 4 R_n^2 R_m^2 \right) 
\nonumber \\ 
&& \left( \sqrt{ (1-R_n^2) (1- R_m^2) - \frac{v^2}{4 }} \right)^{\frac{M}{2}-3}
 \theta \left(v^2 \leq 4 (1-R_n^2) (1- R_m^2) \right) 
\label{pvend}
\end{eqnarray}

For large $M=2^N$, the integral is dominated by the saddle-point
corresponding to $R_n^2=1-R_n^2=\frac{1}{2}=R_m^2=1-R_m^2$,
and one obtains
\begin{eqnarray}
&& {\cal P}(v) \opsimeq_{M \to +\infty} 
 \frac{ \Gamma \left(\frac{M-3}{2}\right)  }{{\sqrt \pi} \Gamma \left(\frac{M}{2}-2\right) }
   \left(  1 - v^2 \right)^{\frac{M}{2}-3}
\label{pvendcol}
\end{eqnarray}
The average of course vanishes by symmetry, and the variance reads
with $M=2^N$
\begin{eqnarray}
<v^2> &&=\int_{-1}^{+1} dv v^2  {\cal P}(v) \opsimeq_{M \to +\infty} \frac{1}{M}
= \frac{1}{2^N}
\label{pvar}
\end{eqnarray}

In the delocalized phase, one thus expect that the off-diagonal Edwards-Anderson order parameter decays as
\begin{eqnarray}
q_{n \ne m}^{deloc}(N) \oppropto_{N \to +\infty}    \frac{1}{2^N}
\label{qoffdelocapp}
\end{eqnarray}

\subsection{ Correlation between magnetizations for two random orthogonal eigenvectors }

Again we consider two random orthogonal eigenvectors satisfying Eqs \ref{norma2}
and Eq \ref{orthog}, but we now focus on the joint distribution
$P(m_n,m_m)$
of the magnetizations on the same site $i=1$
\begin{eqnarray}
m_n && \equiv m_{nn}[i=1] = < \phi_n\vert \sigma_1^z \vert \phi_n>   = \sum_{S_2,..,S_N}  c_n^2 (S_1=+1,S_2..,S_N) - \sum_{S_2,..,S_N}  c_n^2 (S_1=-1,S_2..,S_N) = 2R_n^2-1 
 \\
m_m && \equiv  m_{mm}[i=1] =< \phi_m\vert \sigma_1^z \vert \phi_m>   = \sum_{S_2,..,S_N}  c_m^2 (S_1=+1,S_2..,S_N) - \sum_{S_2,..,S_N}  c_m^2 (S_1=-1,S_2..,S_N) =2 R_m^2-1
\nonumber
\label{mnmm}
\end{eqnarray}
So the joint distribution up to some normalization
can be obtained from the previous computations leading to Eq. \ref{pvend}, by integrating over the variable $v$,
and by inserting the definitions of the two magnetizations via delta functions
\begin{eqnarray}
  P(m_n,m_m) && \propto \int_{-\infty}^{+\infty} dv  \int_0^{1}  R_n dR_n \int_0^{1}  R_m dR_m
\delta (  m_n- (2R_n^2-1) )\delta (  m_m- (2R_m^2-1) )
\nonumber \\ 
&&   \left( \sqrt{ R_n^2 R_m^2 - \frac{v^2}{4 }} \right)^{\frac{M}{2}-3}
 \theta \left(v^2 \leq 4 R_n^2 R_m^2 \right) 
 \left( \sqrt{ (1-R_n^2) (1- R_m^2) - \frac{v^2}{4 }} \right)^{\frac{M}{2}-3}
 \theta \left(v^2 \leq 4 (1-R_n^2) (1- R_m^2) \right) 
\nonumber \\ &&
 \propto  \int_{-\infty}^{+\infty} dv \theta \left(v^2 \leq (1+m_n)(1+m_m) \right) \theta \left(v^2 \leq (1-m_n)(1-m_m) \right) 
\nonumber \\ &&\left( \left[ (1+m_n)(1+m_m)  -v^2 \right]\left[ (1-m_n)(1-m_m)  -v^2 \right] \right)^{\frac{M}{4}-\frac{3}{2}}
\nonumber \\ &&
  \propto \int_{-\infty}^{+\infty} dv \theta \left(v^2 \leq (1+m_n)(1+m_m) \right) \theta \left(v^2 \leq (1-m_n)(1-m_m) \right) 
\nonumber \\ &&e^{\left(\frac{M}{4}-\frac{3}{2}\right) \ln \left[  (1-m_n^2)(1-m_m^2)  - 2v^2  (1+m_n m_m)  +v^4  \right]}
\label{pjointmm}
\end{eqnarray}

For large Hilbert space $M=2^N$, we have already seen that the appropriate rescaled variables for the magnetizations  (Eq \ref {qdiagdelocm2})
and for the variable $v$ (Eq. \ref{pvar}) are
\begin{eqnarray}
m_n && = \frac{\mu_n}{\sqrt{M}} 
\nonumber \\ 
m_m && = \frac{\mu_m}{\sqrt{M}} 
\nonumber \\ 
v && = \frac{w}{\sqrt{M}} 
\label{mumuw}
\end{eqnarray}
Eq. \ref{pjointmm} yields that the joint distribution of the two rescaled magnetizations reads for large $M$
\begin{eqnarray}
  {\cal  P}(\mu_n,\mu_m) && 
  \propto \int_{-\infty}^{+\infty} dw \theta \left(w^2 \leq M \left (1+\frac{\mu_n}{\sqrt{M}} \right)\left (1+\frac{\mu_m}{\sqrt{M}} \right) \right)
\theta \left(w^2 \leq M \left (1-\frac{\mu_n}{\sqrt{M}} \right)\left (1-\frac{\mu_m}{\sqrt{M}} \right) \right)
\nonumber \\ &&e^{\frac{M}{4} \ln \left[  1-\frac{\mu_n^2+ \mu_m^2+2 w^2}{M}   
+ \frac{ (w^2- \mu_n \mu_m  )^2}{ M^2}   \right]}
\nonumber \\ 
 && \propto \int_{-\infty}^{+\infty} dw e^{-\frac{\mu_n^2+ \mu_m^2+2 w^2}{4}   
- \frac{  \mu_n^4+ \mu_m^4+2w^4+4w^2(\mu_n^2+ \mu_m^2+\mu_n\mu_m )  }{ 8M} +o\left( \frac{1}{M} \right)   }
\label{pjointrescal}
\end{eqnarray}
So at leading order, the rescaled magnetizations $\mu_n$ and $\mu_m$ are two independent Gaussian variables
\begin{eqnarray}
  {\cal  P} (\mu_n,\mu_m) && = \frac{1}{4 \pi} e^{-\frac{\mu_n^2+ \mu_m^2}{4} } +O\left( \frac{1}{M} \right)   
\label{pjointrescalinfty}
\end{eqnarray}
The leading contribution to the correlation  
\begin{eqnarray}
 < \mu_n \mu_m > \equiv \int d\mu_n \int d\mu_m  {\cal  P} (\mu_n,\mu_m) \mu_n \mu_m  = - \frac{2}{M}  +o\left( \frac{1}{M} \right)
\label{anticorre}
\end{eqnarray}
is of order $1/M=2^{-N}$ and negative, i.e. the orthogoalization constraint between the two random vectors
leads to a small anticorrelation between
the two magnetizations that vanishes in the thermodynamic limit.

\section{ Properties of the Density correlation matrix for random eigenvectors }

Here we consider two random orthogonal eigenvectors in an Hilbert space of dimension $M=L^d$
i.e. satisfying the normalization conditions
\begin{eqnarray}
 1= \sum_{i=1}^M c_n^2(i)  = \sum_{i=1}^M c_m^2(i) 
\label{norma2bis}
\end{eqnarray}
and the orthogonality condition
\begin{eqnarray}
0= \sum_{i=1}^M c_m(i) c_n(i) 
\label{orthog2}
\end{eqnarray}

We are interested into the joint probability $P(c_n(1),c_m(1))$ of the two first components.
For the $(M-1)$ Cartesian coordinates $c_n(i) $ with $i=2,..,M$
we go to the spherical coordinates of $R_n$.
For the $(M-1)$ Cartesian coordinates $c_m(S_1=+1,..,S_N) $,
we go to the spherical coordinates of radius $R_m$,
and we need to keep the angle $\alpha$ of the scalar product
\begin{eqnarray}
\sum_{i=2}^M c_m (i)  c_n (i) = R_n R_m \cos \alpha
\label{theta2}
\end{eqnarray}

The joint distribution $P(c_n(1),c_m(1))$ reads up to some normalization
\begin{eqnarray}
&& P(c_n(1),c_m(1))  \propto \prod_{i=2}^{M} \left[ \int dc_n(i) \int dc_m(i) \right]
\delta \left( 1- \sum_{i=1}^M c_n^2(i)\right)\delta \left( 1- \sum_{i=1}^M c_m^2(i)\right)
\delta \left( \sum_{i=1}^M c_m (i)  c_n (i)\right)
\nonumber \\
&& \propto 
 \int_0^1 R_n^{M-2} dR_n \int_0^1 R_m^{M-2} dR_m  \int_0^{\pi} d \alpha (\sin \alpha)^{M-3} 
\nonumber \\
&& \delta \left( 1- c_n^2(1)-R_n^2\right)\delta \left( 1-  c_m^2(1) -R_m^2 \right)
\delta \left(  c_m (1)  c_n (1) + R_n R_m \cos \alpha\right)
\nonumber \\
&& \propto 
 \int_0^1 R_n^{M-2} dR_n \int_0^1 R_m^{M-2} dR_m  \int_0^{\pi} d \alpha (\sin \alpha)^{M-3} 
\nonumber \\
&&\frac{\delta \left( R_n-\sqrt{1- c_n^2(1)}\right)}{ 2 R_n }
\frac{ \delta \left( R_m-\sqrt{1-  c_m^2(1) } \right)}{ 2 R_m }
\frac{ \delta \left(  \alpha- \arccos \left(-\frac{c_m (1)  c_n (1)}{R_n R_m}\right) \right)}{ R_n R_m \sin \alpha }
\nonumber \\
&& = \frac{ \frac{M}{2}-1}{\pi}  \left[ 1-c_n^2(1)-c_m^2(1) \right]^{\frac{M}{2}-2 }
\label{pjoint}
\end{eqnarray}
with the joint moments
\begin{eqnarray}
&& \int dc_n(1) dc_m(1) [c_n(1)]^{2k} [c_m(1)]^{2k} P(c_n(1),c_m(1)) 
 = \frac{ \Gamma \left(\frac{M}{2} \right)\Gamma^2 \left(\frac{1}{2} +k \right)}
{ \pi \Gamma \left(\frac{M}{2} +2k \right)} 
\label{pjointckck}
\end{eqnarray}
In particular the value for $k=1$
\begin{eqnarray}
&& \int dc_n(1) dc_m(1) [c_n(1)]^{2} [c_m(1)]^{2} P(c_n(1),c_m(1)) 
 = \frac{ 1}
{  M^2+2M} 
\label{pjointm}
\end{eqnarray}
yields the density correlation
\begin{eqnarray}
\overline{Y_{n \ne m}(M) } && =  M \int dc_n(1) dc_m(1) [c_n(1)]^{2} [c_m(1)]^{2} P(c_n(1),c_m(1)) 
=\frac{ 1} {  M+2} 
\label{z2mdeloc}
\end{eqnarray}

The partial law of the component $c_n(1)$ alone reads
\begin{eqnarray}
&& P_1(c_n(1)) \equiv \int dc_m(1) P(c_n(1),c_m(1)) 
 = \frac{ \Gamma \left(\frac{M}{2} \right)}
{ \sqrt {\pi} \Gamma \left(\frac{M}{2}-\frac{1}{2} \right)}  \left[ 1-c_n^2(1) \right]^{\frac{M}{2}- \frac{3}{2} }
\label{ppartial}
\end{eqnarray}
with the moments
\begin{eqnarray}
&& \int dc_n(1) [c_n(1)]^{2q} P_1(c_n(1))
 = \frac{ \Gamma \left(\frac{M}{2} \right)\Gamma \left(\frac{1}{2} +q \right)}
{ \sqrt {\pi} \Gamma \left(\frac{M}{2} +q \right)} 
\label{ppartialm}
\end{eqnarray}
In particular for $q=1$, one recovers the normalization condition as it should
\begin{eqnarray}
&& \int dc_n(1) [c_n(1)]^{2} P_1(c_n(1))
 = \frac{ 1}{ M} 
\label{ppartialm1}
\end{eqnarray}
The value for $q=2$
\begin{eqnarray}
&& \int dc_n(1) [c_n(1)]^{4} P_1(c_n(1))
 = \frac{ 3}
{  M(M+2)} 
\label{ppartialm2}
\end{eqnarray}
yields the Inverse Participation Ratio 
\begin{eqnarray}
\overline{Y_{diag}(M) } && = M \int dc_n(1) [c_n(1)]^{4} P_1(c_n(1))
 = \frac{ 3}{  (M+2)} 
\label{y2mdeloc}
\end{eqnarray}


\begin{thebibliography}{99}

\bibitem{wigner}
E.P. Wigner, Ann. Math. 53, 36 (1951).

\bibitem{dysonRMT}
F.J. Dyson, J. Math. Phys. 3, 140 (1962).

\bibitem{mehta}
M.L. Mehta, ``Random matrices'', Academic Press (1991).

\bibitem{oxford}
 ``The Oxford Handbook of Random matrix Theory'', 
Ed. G. Akemann, J. Baik and P. Di Francesco,
Oxford University Press (2011).


\bibitem{berry}
M.V. Berry and M. Tabor, Proc. R. Soc. A 356, 375 (1977).

\bibitem{bohigas}
O. Bohigas, M.J. Giannoni and C. Schmidt, Phys. Rev. Lett. 52, 1 (1984).


\bibitem{revue_huse}
R. Nandkishore and D. A. Huse, Ann. Review of Cond. Mat. Phys. 6, 15 (2015).

\bibitem{revue_altman}
 E. Altman and R. Vosk, Ann. Review of Cond. Mat. Phys. 6, 383 (2015).




\bibitem{deutsch}
J.M. Deutsch, Phys. Rev. A 43, 2046 (1991).

\bibitem{srednicki}
M. Srednicki, Phys. Rev. E 50, 888 (1994).

\bibitem{nature}
M. Rigol, V. Dunjko and M. Olshanii, Nature 452, 854 (2008)

\bibitem{mite}
S. Goldstein, D.A. Huse, J.L. Lebowitz and R. Tumulka, Phys. Rev. Lett. 115,
100402 (2015).

\bibitem{rigol}
L. D'Alessio, Y. Kafri, A. Polkovnikov and M. Rigol, arxiv:1509.06411.

\bibitem{emergent_swingle}
B. Swingle, arxiv:1307.0507.

\bibitem{emergent_serbyn}
M. Serbyn, Z. Papic and D.A. Abanin, Phys. Rev. Lett. 111, 127201 (2013).

\bibitem{emergent_huse}
D.A. Huse, R. Nandkishore and V. Oganesyan, Phys. Rev. B 90, 174202 (2014).

\bibitem{emergent_ent}
A. Nanduri, H. Kim and D.A. Huse, Phys. Rev. B 90, 064201 (2014).

\bibitem{imbrie}
J. Z. Imbrie, arxiv:1403.7837.

\bibitem{serbyn_quench}
M. Serbyn, Z. Papic and D.A. Abanin, Phys. Rev. B 90, 174302 (2014).

\bibitem{emergent_vidal}
A. Chandran, I.H. Kim, G. Vidal and  D.A. Abanin, Phys. Rev. B 91, 085425 (2015).

\bibitem{emergent_ros}
V. Ros, M. M\"uller and A. Scardicchio, Nucl. Phys. B 891, 420 (2015).

\bibitem{emergent_rademaker}
L. Rademaker and M. Ortuno, Phys. Rev. Lett. 116, 010404 (2016).

\bibitem{rsrgx}
D. Pekker, G. Refael, E. Altman, E. Demler and V. Oganesyan,
Phys. Rev. X 4, 011052 (2014).

 \bibitem{rsrgx_moore}
Y. Huang and J.E. Moore, Phys. Rev. B 90, 220202(R) (2014).

\bibitem{vasseur_rsrgx}
R. Vasseur, A. C. Potter and S.A. Parameswaran, Phys. Rev. Lett. 114, 217201 (2015).


\bibitem{yang_rsrgx}
M. Pouranvari and K. Yang, arxiv:1508.05829.

\bibitem{rsrgx_bifurcation}
 Y.Z. You, X.L. Qi and C. Xu, arxiv:1508.03635

\bibitem{vasseur_edge}
R. Vasseur, A.J. Friedman, S.A. Parameswaran and A. C. Potter,
arxiv:1510.04282.

\bibitem{c_emergent}
C. Monthus, arxiv:1509.06258.

\bibitem{fisher_AF}
D. S. Fisher, Phys. Rev. B 50, 3799 (1994).

\bibitem{fisher}
D. S. Fisher, Phys. Rev. Lett. 69, 534 (1992); \\
D. S. Fisher, Phys. Rev. B 51, 6411 (1995).

\bibitem{fisherreview}
D. S. Fisher, Physica A 263, 222 (1999).

\bibitem{vosk_rgentanglement}
R. Vosk, D.A. Huse, and E. Altman, Phys. Rev. X 5, 031032 (2015)..

\bibitem{vasseur_resonant}
A. C. Potter, R. Vasseur and S.A. Parameswaran, Phys. Rev. X 5, 031033 (2015).




\bibitem{pal}
A. Pal and D.A. Huse, Phys. Rev. B 82, 174411 (2010).

\bibitem{vadim}
V. Oganesyan and D.A. Huse, Phys. Rev. B 75, 155111 (2007).

\bibitem{alet}
D. J. Luitz, N. Laflorencie and F. Alet, Phys. Rev. B 91, 081103 (2015).

\bibitem{bauer}
B. Bauer and C. Nayak, J. Stat. Mech. P09005 (2013).

\bibitem{kjall}
J. A. Kj\"all, J. H. Bardarson and F. Pollmann, Phys. Rev. Lett. 113, 107204 (2014).

\bibitem{grover}
T. Grover, arxiv:1405.1471.

\bibitem{mirlinrevue}
A.D. Mirlin, Phys. Rep. 326, 259 (2000) ; \\
 F. Evers and A.D. Mirlin, Rev. Mod. Phys. 80, 1355 (2008).

\bibitem{mirlin_evers}
 F. Evers and A. D. Mirlin
Phys. Rev. Lett. 84, 3690 (2000); \\
A.D. Mirlin and F. Evers, 
Phys. Rev. B 62,  7920 (2000).

\bibitem{fyodorov}
Y.V. Fyodorov, A. Ossipov and A. Rodriguez, J. Stat. Mech. L12001 (2009).

\bibitem{fyodorovrigorous}
Y.V. Fyodorov, A. Kupiainen and C. Webb. arxiv:1509.01366.


\bibitem{oleg1}
O. Yevtushenko and V. E. Kratsov,
J. Phys. A 36, 8265 (2003).

\bibitem{oleg2}
O. Yevtushenko and A. Ossipov, 
J. Phys. A 40, 4691 (2007).

\bibitem{oleg3}
S. Kronm\"uller, O. M.  Yevtushenko and E. Cuevas, 
J. Phys. A 43, 075001 (2010).

\bibitem{oleg4}
V. E. Kratsov, A. Ossipov, O. M.  Yevtushenko and E. Cuevas, 
Phys. Rev. B 82, 161102(R) (2010).

\bibitem{us_strongmultif}
C. Monthus and T. Garel, J. Stat. Mech. (2010) P09015.

\bibitem{olivier_per}
E. Bogomolny and O. Giraud, Phys. Rev. E 84, 036212 (2012).

\bibitem{olivier_strong}
E. Bogomolny and O. Giraud, Phys. Rev. E 84, 046208 (2012).

\bibitem{olivier_conjecture}
E. Bogomolny and O. Giraud, Phys. Rev. Lett. 106, 044101 (2011).


\bibitem{c_reso}
C. Monthus, arxiv:1510.03711.








\bibitem{dyson}
F.J. Dyson, J. Math. Phys. 3, 1191 (1962).


\bibitem{chalker}
J.T. Chalker, I.V. Lerner, and R.A. Smith, Phys. Rev. Lett. 77, 554 (1996).

\bibitem{serbyn}
M. Serbyn and J.E. Moore, arxiv:1508.07293.

\bibitem{papic}
M. Serbyn, Z. Papic and D.A. Abanin, arxiv:1507.01635.

\bibitem{gardiner}
C. W. Gardiner, ``Handbook of Stochastic Methods: for Physics, Chemistry and the Natural Sciences'' (Springer Series in Synergetics), Berlin (1985).

\bibitem{vankampen}
N.G. Van Kampen, ``Stochastic processes in physics and chemistry'', Elsevier Amsterdam (1992).

\bibitem{risken}
H. Risken, ``The Fokker-Planck equation : methods of solutions and applications'', Springer Verlag Berlin (1989).

\bibitem{EA}
S.F. Edwards and P.W. Anderson, J. Phys. F Met. Phys. 5, 965, (1975).

\bibitem{louck}
J.D. Louck, Foundations of Physics 27, 8 (1997).

\bibitem{luck}
J.M. Luck, arxiv:1510.06163.

\bibitem{Ludwig}
A.W.W. Ludwig, Nucl. Phys. B 330, 639 (1990).


\bibitem{Jac_Car}
J.L. Jacobsen and J.L. Cardy, Nucl. Phys., B515, 701 (1998).

\bibitem{Ols_You}
T. Olsson and A.P. Young, Phys. Rev., B60, 3428 (1999).

\bibitem{Cha_Ber}
C. Chatelain and B. Berche, Nucl. Phys., B572, 626 (2000).

\bibitem{Cha_Ber_Sh}
C. Chatelain, B. Berche and L.N. Shchur, J. Phys. A Math. Gen. 34, 9593 (2001).

\bibitem{PCBI}
G. Pal\'agyi, C. Chatelain, B, Berche and F. Igl\'oi, Eur. Phys. J.
B13, 357 (2000).


\bibitem{BCrevue}
B. Berche and C. Chatelain, in {\it Order, disorder, and criticality}, ed. 
by Yu. Holovatch, World Scientific, Singapore 2004, p.146. 

\bibitem{Sourlas}
N. Sourlas, Europhys. Lett. 3, 1007 (1987).

\bibitem{Thi_Hil}
M.J. Thill and H.J. Hilhorst, J. Phys. I 6, 67 (1996).

\bibitem{Par_Sou}
G. Parisi and N. Sourlas, Phys. Rev. Lett. 89, 257204 (2002).

\bibitem{DPmultif}
C. Monthus and T. Garel, Phys. Rev. E 75, 051122 (2007).

\bibitem{wettingmultif}
C. Monthus and T. Garel, Phys. Rev. E 76, 021114 (2007). 

\bibitem{nancy}
C. Monthus, B. Berche and C. Chatelain, J. Stat. Mech. P12002 (2009).

\bibitem{plasma}
A.G. Aronov, V.E. Kravtsov and I.V. Lerner, JETP Lett. 59, 39 (1994) ; \\
 V.E. Kravtsov and I.V. Lerner, J. Phys. A Math. Gen. 28, 3623 (1995).

\bibitem{conjectured2}
J. T. Chalker, V.E. Kravtsov and I.V. Lerner, JETP Lett. 64, 386 (1996).


 \end{thebibliography}
\end{document}